\documentclass[10pt,preprint]{aastex}
\usepackage{graphicx}
 \usepackage{epstopdf}
\usepackage{epsfig}
\usepackage{amsmath}
\usepackage{multirow}
\usepackage{lineno}
 \usepackage{color}
\begin{document}

\title{Particle acceleration in mildly--relativistic shearing flows: the interplay of systematic and stochastic effects, 
       and the origin of the extended high-energy emission in AGN jets}
\author{Ruo-Yu Liu$^{1}$, F.M. Rieger$^{1,2}$ and F.A. Aharonian$^{1,3}$}
\altaffiltext{1}{Max-Planck-Institut f\"ur Kernphysik, Saupfercheckweg 1, 69117 Heidelberg, Germany; ruoyu@mpi-hd.mpg.de; frank.rieger@mpi-hd.mpg.de; aharon@mpi-hd.mpg.de}
\altaffiltext{2}{ZAH, Institut f\"ur Theoretische Astrophysik, Universit\"at Heidelberg, 
       Philosophenweg 12, 69120 Heidelberg, Germany}
\altaffiltext{3}{Dublin Institute for Advanced Studies, 31 Fitzwilliam Place, Dublin 2, Ireland}

\begin{abstract}
The origin of the extended X-ray emission in the large-scale jets of active galactic nuclei (AGNs) poses challenges to conventional models of acceleration and 
emission. {Although the electron synchrotron radiation is considered the most feasible radiation mechanism, the formation of the continuous large-scale X-ray structure remains an open issue.}
As astrophysical jets are expected to exhibit some turbulence and shearing motion, we here investigate the potential of shearing flows to facilitate an 
extended acceleration of particles and evaluate its impact on the resultant particle distribution. Our treatment incorporates systematic shear and 
stochastic second-order Fermi effects. We show that for typical parameters applicable to large-scale AGN jets, stochastic second-order Fermi 
acceleration, which always accompanies shear particle acceleration, can play an important role in facilitating the whole process of particle energization. 
We study the time-dependent evolution of the resultant particle distribution in the presence of second-order Fermi acceleration, shear acceleration, and 
synchrotron losses using a simple Fokker--Planck approach and provide illustrations for the possible emergence of a complex (multicomponent) 
particle energy distribution with different spectral branches. We present examples for typical parameters applicable to large-scale AGN jets, indicating 
the relevance of the underlying processes for understanding the extended X-ray emission and the origin of ultrahigh-energy cosmic rays.
\end{abstract}

\section{Introduction}
The detection of extended X-ray emission from the kilo-to-megaparsec-scale jets in Active Galactic Nuclei (AGNs), as seen, e.g. in 
3C~15 \citep{Kataoka03}, 3C~120 \citep{Harris04}, 3C~273 \citep{Jester06}, Cen~A \citep{Kraft02,Kataoka06}, M87 \citep{Wilson02, 
Marshall02}, Pictor A \citep{Wilson01,Hardcastle05}, PKS~1127-145 \citep{Siemiginowska02}, PKS~1136-135 \citep{Sambruna06} 
and PKS~0637-752 \citep{Chartas00}, has raised questions about the origin of this emission and the efficiency of particle acceleration 
operating in these jets. Shock acceleration, in particular, is an efficient process fro particle energization and thought to be responsible 
for many nonthermal high-energy phenomena in the universe \citep[e.g.][]{Bell13,Lemoine14,Marcowith16}. Given typical equipartition 
magnetic field strengths of $10-100\,\mu$G in the jets, $\sim 100\,$TeV electrons would be required to produce X-rays via synchrotron 
radiation. While strong shocks may be powerful enough to accelerate electrons to such high energies, the associated synchrotron cooling 
time is only a few thousand years. Hence, even if these nonthermal electrons are taken to propagate quasi-rectilinearly, their energies 
would be exhausted before going beyond some hundreds of parsecs, so that one would not expect to see an extended morphology of 
these jets in X-rays. However, the projected lengths of some of these X-ray jets extend to $\sim (10-100)$\,kpc, such as those in 3C~273, 
PKS~0637-752 and PKS~1127-145, not to mention their physical lengths. Even the knots and hotspots in these jets sometimes have 
projected sizes of more than one kiloparsec, which seems challenging to account for if the X-ray-emitting electrons are only taken to 
be accelerated (quasi-locally) at some shock front and not re-energized by some distributed mechanisms. Alternative scenarios where 
the X-ray emission is not related to electron synchrotron radiation are in principle conceivable. Proton synchrotron radiation of $>10^{18}
\,$eV protons \citep{Aharonian02,Bhattacharyya16}, for example, could offer a convenient way to account for the extended, large-scale 
X-ray emission, but it tends to need rather high magnetic fields, typically more than 100\,$\mu$G. Inverse Compton (IC) models, in which 
electrons with Lorentz factor $\gamma\sim 100$ in the rest frame of a highly relativistic jet (with bulk Lorentz factors of $\sim 10-20$ out 
to kpc-Mpc scales) upscatter cosmic microwave background (CMB) photons to the X-ray regime \citep{Tavecchio00,
Celotti01,Ghisellini05}, do not call for some distributed (re)acceleration since the lifetimes of the electrons responsible for 
the X-rays via IC are sufficiently long. However, these IC models require jets that are well aligned with the line of sight and thereby tend to 
imply rather uncomfortably large physical jets lengths, sometimes in excess of $\sim$Mpc. Furthermore, the IC jets would need to maintain 
their highly relativistic speeds up to these scales. It has been argued recently that current polarimetric results and high-energy gamma-ray 
constraints are providing corroborating evidence for disfavoring IC/CMB models \citep[e.g.,][but see also Lucchini et al. 2016]{Cara13, Meyer15, Meyer17}. 
Thus, if one wants to hold on to the idea that the X-ray emission is caused by electron synchrotron radiation, 
some spatially distributed (re)acceleration along the large-scale jet is needed. In principle, stochastic second-order Fermi 
acceleration could be a potential candidate for such a distributed mechanism\cite{Stawarz02}, provided that high enough Alfv\'{e}n speeds $v_A$ would be 
occurring in the large-scale jets of AGN ($t_{\rm acc} \sim (c/v_A)^2 \tau$, $\tau$ the mean scattering time). In a complementary way, if a velocity
gradient exists within the flow, particles traversing it could sample the flow difference and undergo shear particle acceleration. This could
potentially facilitate the energization of particles such as to reproduce the observed extended high-energy emission \cite[e.g.,][]{Rieger07}. 
When jets interact with the ambient medium, their boundary interface could in fact experience significant deceleration, while the 
central parts may remain less affected and still sustain relatively high bulk flow velocities. Hence a longitudinal shear structure might naturally 
arise across a jet. There is mounting phenomenological evidence for the existence of such velocity shear structures in different astrophysical 
jet sources \cite[e.g.,][]{Laing14,Lundman14}. Efficient shear acceleration in such flows, however, usually depends on the injection of 
energetic seed particles and thus needs some effective catalyst for its operation. The current paper focuses on the (re-)energization of 
particles and the possible interplay between stochastic and shear effects. 
Our main purpose is to provide an analysis of shear acceleration in a mildly relativistic flow for parameters appropriate to the large-scale
jets in AGNs and to analyse the contribution of the systematic (associated with shear) and stochastic (associated with classical second-order 
Fermi) velocity components of the scattering center's motion in the acceleration process. {  While we are particularly interested in the 
acceleration of electrons and their resultant synchrotron radiation, shear acceleration might also be of relevance for the acceleration of
protons to ultrahigh energies (UHEs).} Possible astrophysical sites for UHE cosmic-ray acceleration are limited, though large-scale AGN 
jets might still be counted amongst the most promising ones \cite[e.g.][]{Hillas84, Aharonian02b}. For this reason we also examine the 
relevance of our framework for the possible energization of UHE cosmic rays.

The paper is organized as follows. In Section 2, the main aspects of shear acceleration are briefly described based on a heuristic microscopic 
treatment, and applied to the acceleration process in a cylindrically symmetric, relativistic jet of gradual shear. In Section 3, we study the 
time evolution of the spectrum of the accelerated particles in a shearing flow within a Fokker--Planck approach, with a particular focus on 
the effect of stochastic acceleration, and examine its implications for the high-energy emission from large-scale AGN jets. Section 4 explores 
the potential of shearing jets for the acceleration of UHE cosmic rays. Discussions and conclusions are presented in Section 5.

\section{Shear Particle Acceleration}
\subsection{Conceptual Approaches and Applications}
It has been recognized since the 1980s that charged particles could undergo acceleration in shearing outflows. Energetic particles could 
sample the velocity difference while they move across the flow by scattering off magnetic field inhomogeneities that are embedded in 
different layers of the shearing flow with different bulk velocities. This is accompanied by the conversion of bulk motion kinetic energy of the 
background flows to nonthermal particle energies. {  In certain respects, shear particle acceleration could be conceived of as a Fermi-type 
mechanism, showing some resemblance to shock or stochastic acceleration \citep{Rieger07}, though with a different kinetic origin related 
to the systematic component of the scattering centers' motion.}

In the pioneering work of Berezhko and collaborators \citep{Berezhko81a,Berezhko81b, Berezhko82}, particle acceleration in a nonrelativistic gradual shearing flow was considered, and steady-state particle distributions in momentum space were studied based on an analysis of the 
relevant nonrelativistic Boltzmann equation. An independent, seminal study was performed later on by \citet{Earl88} by expanding the Boltzmann 
equation to the lowest order in the ratio of flow speed to random particle speed, and obtaining Parker's equation augmented by terms describing 
cosmic ray viscosity and inertial drifts. Extensions to discontinuous or non-gradual shearing flows were studied by \citet{Jokipii89}, while extensions 
to the relativistic regime of gradual shearing flows were obtained by \citet{Webb89} and later on applied to characteristic flow profiles \citep{Rieger02,
Rieger04,Rieger16}. 

A complementary "microscopic" approach to study particle energization in gradual shearing flows instead of the traditional method of solving the
Boltzmann equation was presented in \citet{Jokipii90}. They assumed an energy--independent mean free path of the particles undergoing 
scattering, and derived the average momentum change and dispersion per mean free path, i.e., the acceleration rate and the momentum 
diffusion rate. Extensions to the more general case of an energy--dependent particle mean free path and related applications have been 
presented in \citet{Rieger06}.

Monte Carlo simulations to study the spectrum shape of the accelerated particles in relativistic non-gradual shearing flows were performed by 
\citet{Ostrowski90}. These and related results led to the suggestion that (ultra)relativistic shearing flows might be potential sources of UHE cosmic-ray protons \citep{Ostrowski98, Ostrowski00,Rieger04}. 
 
As shown in previous studies, the efficiency of shear acceleration is closely related to the issue of how strong the velocity shear is, and hence it operates
more efficiently in relativistic jets \citep{Webb89,Ostrowski90,Rieger04}. An unusual feature of particle acceleration in gradual shearing flow is 
that its efficiency increases with increasing particle mean free path (i.e., with particle's energy), which is in contrast to both the (first-order Fermi) 
shock acceleration and classical stochastic (second-order Fermi) acceleration. This is related to the fact that more energetic particles tend to 
have larger mean free paths and therefore are capable to experience a larger velocity difference of background flow before being scattered 
again. This in turn allows for a higher relative velocity of the scattering centers and hence leads to a larger energy increase after scattering
\citep[][]{Rieger06}. As a consequence, however, this mechanism tends to be very inefficient for the acceleration of low energy particles. 
Therefore, for gradual shear acceleration to start operating efficiently, an injection of high-energy seed particles is needed. In the current study 
the role of stochastic second-order Fermi acceleration in this context is evaluated.

\subsection{A simplified microscopic analysis of gradual shear acceleration}
In this subsection, we briefly introduce the shear acceleration mechanism by analysing the associated diffusion rate of particles in momentum 
space and the average rate of change of energy (i.e., the Fokker--Planck coefficients). To make the underlying physical picture more transparent, 
we employ a simplified microscopic treatment similar to that adopted in \citet{Jokipii90} and \cite{Rieger06} to derive the Fokker--Planck 
coefficients, and apply it to a relativistic 3D shearing jets with cylindrical symmetry. 

\begin{figure}
\includegraphics[width=0.6\textwidth]{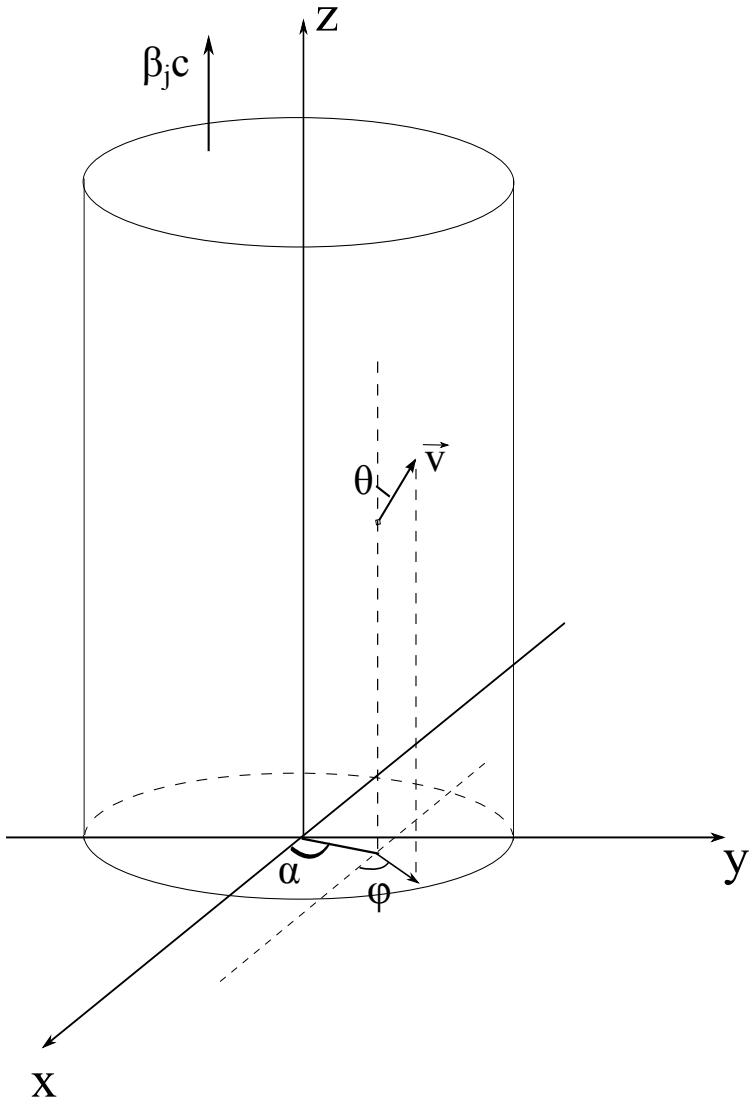} \caption{Sketch of the assumed geometrical setup of the jet.\label{sketch}}
\end{figure}

Consider a cylindrical shearing jet propagating along the z-axis with a transverse velocity profile that decreases with jet radius $r$ (or the perpendicular 
distance to the $z$-axis), as shown in Fig.~\ref{sketch}. Given the cylindrical symmetry, the jet can be regarded as a collection of a series of coaxial 
cylindrical layers with different radii, and the bulk velocity of each layer can be denoted by $\beta_j(r)\,c$, with  $r=\sqrt{x^2+y^2}$ and $\beta_j(r)$ 
being the velocity of the layer at $r$ in units of the light speed $c$. 

Now assume that a charged particle (e.g., electron or proton) of mass $m$ at $(x,y,z)$ with momentum $p=\gamma \beta mc$ is moving towards a 
certain direction, where $\gamma$ and $\beta c$ are the Lorentz factor and the velocity of the particle, respectively. Denote the angle between the 
particle's velocity and $z$-axis by $\theta$, the angle between the projection of the particle's velocity on $xy$-plane and $x$-axis by $\phi$, we can 
write the velocity components in $x$, $y$, and $z$ directions as $\beta_{x}=\beta{\rm sin} \theta{\rm cos}\phi$, $\beta_{y}=\beta{\rm sin} \theta{\rm sin}\phi$ and 
$\beta_{z}=\beta{\rm cos} \theta$ respectively.  In a scattering event, the particle's momentum direction is considered to be randomized and its 
magnitude conserved in the local flow frame. Between two scattering events it moves across the flow, and we use $\tau$ and $\lambda$ to denote 
the available mean scattering time and particle mean path, respectively. Shear flows are likely to excite turbulence, and their diffusion characteristics 
are not well known. For simplicity we parameterize here the spatial diffusion of particles by using a prescription from quasi-linear theory ($\delta B \ll B$).
Accordingly, the mean free path of a charged particle is written as\footnote{Numerical simulation shows that the mean free path of a charged particle  in the presence of strong turbulence keeps the same energy dependence as that in the quasi-linear theory \cite{Casse02}.}
\begin{equation}
\lambda\simeq \frac{B_0^2/8\pi}{k_0I(k_0)}r_g\left(\frac{r_g}{\Lambda_{\rm max}}\right)^{1-q}= \xi^{-1} r_g\left(\frac{r_g}{\Lambda_{\rm max}}\right)^{1-q}
\propto \gamma^{2-q}\,,
\end{equation} 
where $B_0$ is the strength of the regular magnetic field in the jet assumed to be directed along the jet axis, and $r_g=1.7\times 10^{16}\,{\rm cm}
~(\gamma/10^8)(B_0/10\mu G)^{-1}$ is the Larmor radius of the particle. $I(k)$ is the magnetic energy density per unit wavenumber $k$ in the turbulent 
magnetic field. The turbulent energy spectrum is assumed to be a power law of index $q$, i.e. $I(k)\propto k^{-q}$. $\xi\equiv \frac{k_0 I(k_0)}{B_0^2/8\pi}$ 
is defined as the ratio of energy density between the turbulent magnetic field and the regular magnetic field. $\Lambda_{\rm max}$ is the longest interacting wavelength of the turbulence. 

In such a treatment, the particle will move to $(x+\Delta x, y+\Delta y, z+\Delta z)$ within one mean free path, where
\begin{equation}\label{dxdydz}
\left\{
\begin{array}{lll}
\Delta x=\beta_{x}c\tau(p)\\
\Delta y=\beta_{y}c\tau(p)\\
\Delta z=\beta_{z}c\tau(p),
\end{array}
\right.
\end{equation} 
with $\tau(p)\equiv \lambda(p)/c$ being the local scattering time. We can then express the transverse distance that a particle crosses (which 
in turn determines the relative velocity difference between the two layers) by
\begin{equation}\label{dr}
\Delta r=\sqrt{(r{\rm cos}\alpha +\Delta x)^2+(r{\rm sin}\alpha +\Delta y)^2}-r
\end{equation}
where $\alpha={\rm tan}^{-1}(y/x)$. Substituting Eq.(\ref{dxdydz}) into Eq.(\ref{dr}), we obtain
\begin{equation}\label{dr2}
\Delta r=\left[r^2+2\beta cr\tau(p)\sin\theta {\rm cos}(\alpha-\phi)+\beta^2c^2{\rm sin}^2\theta\tau(p)^2\right]^{1/2}-r.
\end{equation}
Denoting the velocity of the layer at radius $r+\Delta r$ by $\beta_{j}^{'}$, the relative velocity and Lorentz factor of the layer at radius $r+\Delta r$ with 
respect to that of the layer at radius $r$ can be given by $\beta_\Delta=\frac{\beta_{j}^{'}-\beta_{j}}{1-\beta_{j}\beta_{j}'}$ and $\Gamma_\Delta=
(1-\beta_\Delta^2)^{-1/2}$ respectively.  Once we know the shear profile of the jet and $\Delta r$, we can obtain the velocity or Lorentz factor of the 
particle in rest frame of the new layer by 
\begin{equation}\label{gp}
\gamma '=\Gamma_{\Delta}\gamma(1-\beta\beta_\Delta {\rm cos}\theta).
\end{equation}
{In the context of this work, the particle distributions are assumed to be quasi-homogeneous and isotropic in both real and momentum space. 
We also assume in the following that the diffusion of particles is approximately isotropic. This implicitly presumes the presence of strong turbulence 
($\Delta B\sim B$) to facilitate efficient cross-field diffusion. While these are certainly strong assumptions and more complex situations clearly need 
to be studied to achieve a fuller picture, these descriptions allow for a transparent analytical treatment and restrict the number of free parameters, 
thus providing a useful starting point. } 
 
Taking these assumptions into account, we can average over $\phi$, $\theta$ and $\alpha$ to get the mean energy increase in one scattering. 
From Eq.~(\ref{dr2}) one finds that for any given $\phi$ we can always redefine $\alpha=\alpha-\phi$ so that as long as we take the average of both 
$\alpha$ and $\phi$ from 0 to $2\pi$, the result will not change. Thus, for simplicity, hereafter we fix $\phi=0$. 

We focus on a nonrelativistic gradual shear flow in which the velocity difference between two layers is much less than the speed of light. Note 
that this could be a common case in mildly relativistic large-scale jets (e.g., jets of radio galaxies). Thus, we may 
expect $\vert\Gamma_{j}^2\Delta \beta_j \vert \ll 1$, which gives
\begin{equation}\label{bdelta}
\beta_{\Delta}=\frac{\beta_j'-\beta_j}{1-\beta_j'\beta_j}=\frac{\Gamma_j^2 \Delta \beta_j}{1-\Gamma_j^2\Delta \beta_j \beta_{j}}
                   \simeq \Gamma_j^2\Delta\beta_j\,(1+\Gamma_j^2\Delta\beta_j\,\beta_j)
\end{equation}
and 
\begin{equation}\label{gdelta}
\Gamma_{\Delta}=\frac{1}{\sqrt{1-\beta_\Delta^2}}\simeq 1+\frac{1}{2}\beta_\Delta^2\,.
\end{equation}
In the above two equations, $\Delta\beta_j \equiv \beta_{j}'-\beta_{j}\simeq \frac{\partial \beta_j}{\partial r}\Delta r$. Given that $\Delta r \ll r$, 
Eq.~(\ref{dr}) can be reduced to $\Delta r \simeq \Delta x {\rm cos}\alpha$ to the first order.  Thus Eq.~(\ref{gp}) can be written in the form
\begin{equation}
\gamma '\simeq\gamma(1+\frac{1}{2}\beta_\Delta^2-\beta\beta_\Delta{\rm cos}\theta)
\end{equation} 
when keeping the expression to the second order. Substituting Eqs.~(\ref{bdelta}) and (\ref{gdelta}) into the above equations, we obtain
\begin{equation}\label{newgamma}
\gamma ' \simeq\gamma(1+\frac{1}{2}A^2\beta^2\tau^2{\rm sin}^2\theta{\rm cos}^2\alpha- 
                 A\beta^2\tau{\rm sin}\theta{\rm cos}\theta{\rm cos}\alpha - A^2\beta_j\beta^3\tau{\rm sin}^2\theta{\rm cos}\theta{\rm cos^2}\alpha)
\end{equation}
where $A\equiv \Gamma_j^2(\frac{\partial \beta_j}{\partial r})c$ depends only on the shear profile of the jet.
The average rate of momentum dispersion, which is related to the diffusion coefficient in momentum (particle Lorentz factor) space, might then 
be calculated as
\begin{equation}\label{fpcoef2}
\langle \frac{\Delta \gamma^2}{\Delta t} \rangle \equiv \frac{2\langle (\gamma '-\gamma)^2\rangle }{\tau}
=2\,\int\int\int (\gamma '-\gamma)^2{\rm sin}\theta d\theta\ d\phi d\alpha/(8\pi^2 \tau)
\simeq \frac{2}{15}A^2\gamma^2\tau\propto \gamma^{4-q}\,.
\end{equation}
In the last expression we have dropped the dependence on the particle velocity since we focus on ultrarelativistic particles whose $\beta 
\rightarrow 1$. Expression Eq.~(\ref{fpcoef2}) is also known as Fokker--Planck diffusion coefficient\footnote{The original Fokker--Planck 
coefficients are defined as $\langle \Delta p^2/\Delta t \rangle$ and $\langle \Delta p/\Delta t \rangle$. But given the linear relation 
between the Lorentz factor and the momentum for relativistic particles, we here consider the Fokker--Planck coefficients for the Lorentz 
factor.}. Under the condition of detailed balance \citep{Rieger06}, the first Fokker--Planck coefficient, describing the average rate of 
momentum change, can then be obtained from
\begin{equation}\label{fpcoef1}
\langle\frac{\Delta \gamma}{\Delta t} \rangle= \frac{1}{2\gamma^2}\frac{\partial}{\partial \gamma}\left[\gamma^2 
\langle \frac{\Delta \gamma^2}{\Delta t}\rangle \right]=\frac{6-q}{15}A^2\gamma\tau\propto \gamma^{3-q}\,.
\end{equation}
These expressions coincide with those for non-relativistic shearing jets \cite[see e.g.][]{Rieger06} if $\Gamma_j \to 1$, which qualifies
the derivation here. Note that for the chosen geometrical setup, $\Delta r \gg r$ might in principle occur for high-energy particles, in 
which case we would have $\Delta r\simeq \Delta x$ and the dependence on $\alpha$ in the expression for $\Delta r$ (Eq.~\ref{dr}) 
could be dropped. Provided the velocity profile would be purely linearly decreasing, repeating the same calculations above would then 
formally yield $\langle \Delta \gamma^2/\Delta t \rangle=\frac{4}{15} A^2\gamma^2\tau$, i.e. twice the expression in  Eq.~(\ref{fpcoef2}), as 
long as $\Gamma_{j}^2\Delta \beta_j \ll 1$ applies and deviations from local isotropy are neglected. Similar to the situation for relativistic 
shocks, the latter case may, however, not be realized.

\begin{figure}
\includegraphics[width=0.8\textwidth]{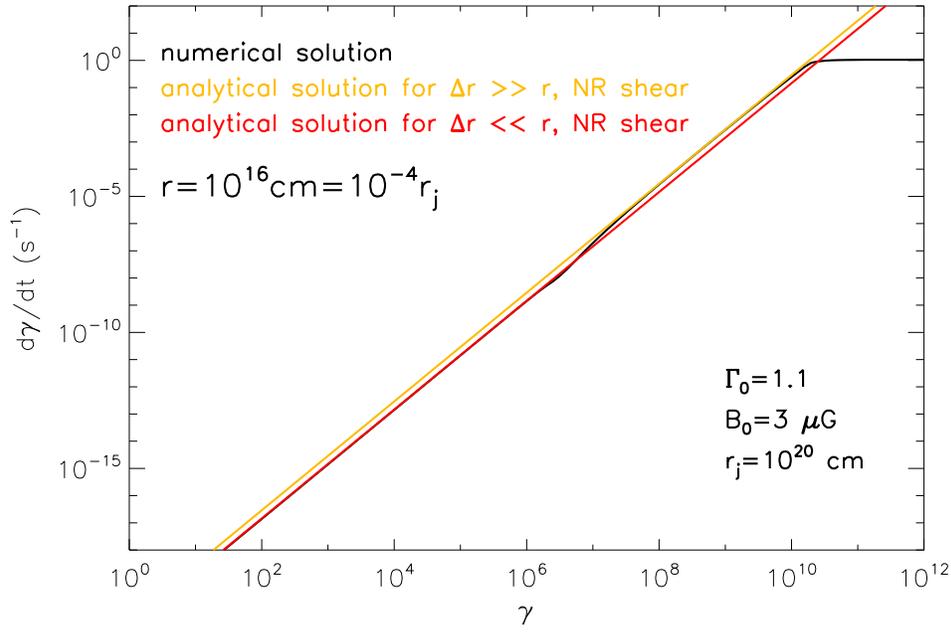}\caption{Illustration of the {nominal} average acceleration rate of electrons starting at position $r$ in a 
large-scale shearing flow, in which the bulk outflow velocity decreases linearly with radial distance from the axis. {For the considered parameters,
the acceleration rate at low $\gamma$ is very slow, illustrating the need for some additional pre-acceleration mechanism.}
 \label{fig:acc_E&R}}
\end{figure}

Fig.~\ref{fig:acc_E&R} shows a numerical example for the (nominal) average shear acceleration rate of electrons versus energies (Lorentz factors),
$\langle \Delta \gamma/\Delta t\rangle$ for electrons starting at $r=10^{16}\rm cm$ in a large-scale nonrelativistic flow with a transverse 
jet radius of $r_j=10^{20}\rm cm$ and bulk Lorentz factor of $\Gamma_{j,0}=1.1$. The velocity of the jet is assumed to decrease linearly with 
the distance $r$ from the axis, from $\beta_{j,0}=\sqrt{1-1/\Gamma_{j,0}^2}$ at $r=0$ to $\beta_{j}=0$ at $r=r_j$ (i.e., $\Delta L=r_j=10^{20}\rm 
cm$). The average magnetic field is considered to be $B=3\mu G$, while $\xi=0.1$, $q=1$, and $\Lambda=10^{18}\rm cm$ are taken for the 
turbulent magnetic field. One can see that $\langle \Delta \gamma/\Delta t\rangle \propto \gamma^{2}$ {which is identical to that given by the analytical expression (Eq.~\ref{fpcoef1}).}  The curve of $\langle \Delta \gamma/
\Delta t\rangle$ becomes flat above $\gamma=2\times 10^{10}$. This is because the mean free path of these electrons is longer than the transverse size 
of the jet so they escape the jet before being scattered. The energy change is then simply due to the Lorentz transformation from the jet frame 
with $\Gamma=\Gamma_j$ to the observer's frame with $\Gamma=1$. So $\Delta \gamma/\Delta t=(\Gamma_{j,0}-1)\gamma/\tau \propto 
\gamma^0$ in the case $q=1$. 
Note, however, that particles radially escaping the jet are unlikely to return, so that particles with energies in such a range are not expected to be
further accelerated. 

\noindent
The timescale for shear acceleration can in principle be estimated from
\begin{equation}\label{ts_non}
t_{\rm acc, sh}=\frac{\gamma}{\langle\Delta\gamma/\Delta t\rangle}=\frac{15}{6-q}A^{-2}\tau^{-1} \propto \gamma^{q-2}
\end{equation}
There is a possibility that particles might escape as they are diffusing inside the jet. The spatial diffusion coefficient is usually related to the particle's 
mean free path $\lambda$ as $\kappa\simeq c\lambda/3$, so that the timescale for diffusive escape might be estimated by 
\begin{equation}
t_{\rm esc}=r_j^2/\kappa=\frac{3r_j^2}{c^2}\tau^{-1}.
\end{equation}
Note that the ratio between the timescale for shear acceleration  and that for diffusive escape then does not depend on the energy of the particle. 
Particles can experience shear acceleration when the escape timescale is longer than the acceleration timescale ($t_{\rm esc}>t_{\rm acc}$). This 
condition can be satisfied when
\begin{equation}\label{con_esc}
\frac{6-q}{5}\Gamma_j^4\left(\frac{\beta_{j,0}}{\Delta L} \right)^2r_j^2 > 1\,,
\end{equation} where $\Delta L$ is the typical radial length scale characterizing the decrease in bulk flow velocity.

Using Eq.~(\ref{ts_non}) and noting that $A\tau \simeq \Gamma_j^2\frac{\partial \beta_j}{\partial r}\tau c \simeq \beta_\Delta$ for a 
mildly relativistic jet, the timescale for gradual shear acceleration could also be written as
\begin{equation}
t_{\rm acc, sh}=\eta \frac{r_g}{c} \beta_\Delta^{-2}\,
\end{equation}
where $\eta=\frac{15}{6-q}\left(\frac{r_g}{\Lambda_{\rm max}}\right)^{1-q}\xi^{-1}$. One can see that the timescale for shear acceleration 
has the same form as that for (nonrelativistic) diffusive shock acceleration or stochastic (second-order Fermi) acceleration, but with 
$\beta_{\Delta}$ the characteristic velocity difference of the background flow experienced by the particle between two scattering events, 
instead of the shock speed of the MHD wave (e.g. Alfv\'{e}n wave), speed which are the relevant velocities in the latter two mechanisms, 
respectively. Indeed, all these three mechanisms draw on a scattering process to convert kinetic energy of the bulk or scattering centers' 
motion to particles. Compared to the classical second-order Fermi acceleration, shear acceleration could achieve higher particle energy 
in environments where the flow speed is larger than the Alfv\'{e}n speed, as expected for the noted large-scale jets in AGNs. When 
compared with shock acceleration, particles could easily enter another flow layer from the side whatever the flow speed is, while in the
shock acceleration mechanism particles may have difficulty in catching up with the shock if the shock's speed is relativistic \citep[unless specific conditions 
prevail, see e.g.][]{Derishev03, Lemoine10}. Unlike the shock and/or stochastic second-order Fermi acceleration, the timescale for shear acceleration 
for $q<2$ is inversely proportional to the energy of the particle, i.e., higher energy particles will get accelerated more easily. This is explained 
by the fact that more energetic particles have longer mean free paths and hence experience a larger velocity difference of the background 
flow between two scatterings. Although this effect favors the acceleration of high energy particles, it also means that the gradual shear 
acceleration is inefficient for low-energy particles. Hence, high-energy "seed'' particles are needed for efficient shear acceleration 
\citep[e.g.,][]{Rieger06}. These energetic seed particles could be provided by other acceleration mechanisms. 
Here we explore the suggestion that for the large-scale AGN jets stochastic (second-order Fermi) acceleration might be a natural supplier 
for such seed particles. In principle, stochastic acceleration is always likely to be present in a general background flow. In the following 
section, we  study the time--dependent spectrum of accelerated particles in a mildly--relativistic shearing flow, paying particular attention to 
the contribution of stochastic acceleration in the process.

\section{The energy distribution of accelerated particles}
The evolution of the particle distribution in momentum space $f(p,t)$ for both stochastic and shear acceleration mechanism can be described by 
a Fokker--Planck-type equation \citep[cf.][]{Rieger07}
\begin{equation}\label{fp_eq_p}
\frac{\partial f(p,t)}{\partial t}=\frac{1}{2p^2}\frac{\partial ^2}{\partial p^2}\left[ p^2\left \langle \frac{\Delta p^2}{\Delta t}\right\rangle f(p,t) \right]
-\frac{1}{p^2}\frac{\partial}{\partial p}\left[p^2\left(\left\langle \frac{\Delta p}{\Delta t}\right\rangle 
+\langle \dot{p}_c\rangle\right)f(p,t)\right]-\frac{f(p,t)}{t_{\rm esc}(p)}+q(p,t)\,.
\end{equation}
Here, the first term on the right-hand side describes diffusion in momentum space. The second term accounts for the systematic energy 
change of particles, including particle acceleration and cooling, which is denoted by $\langle \dot{p}_c\rangle$. The third term considers 
a simplified diffusive escape of particles from the accelerated region by assuming that all particles have finite escape possibilities that 
do not depend on their positions in the acceleration region. In the numerical solution, we set the Fokker-Planck coefficients to be zero 
when the mean free path of a particle becomes larger than its distance to the boundary of the jet, as such particles are likely to escape 
quickly from it. The particle injection term $q(p,t)$ is assumed to be a $\delta$-function, i.e., $q \propto \delta(p-p_0)$ where $p_0$ is the 
injection momentum. In the numerical calculation we will approximate this term as a narrow Gaussian function. Given that $p\simeq \gamma m c$  for ultrarelativistic particles and $n(\gamma,t)d\gamma=4\pi p^2f(p,t)dp$ for an isotropic distribution of particles in momentum space, 
Eq.~(\ref{fp_eq_p}) can be rewritten in the form
\begin{equation}\label{fp_eq}
\frac{\partial n(\gamma,t)}{\partial t}=\frac{1}{2}\frac{\partial}{\partial \gamma}\left[\left \langle \frac{\Delta \gamma^2}{\Delta t}\right\rangle\frac{\partial n(\gamma,t)}{\partial \gamma} \right]- \frac{\partial}{\partial \gamma}\left[ \left( \left\langle \frac{\Delta \gamma}{\Delta t}\right\rangle-\frac{1}{2}\frac{\partial}{\partial \gamma}\left\langle \frac{\Delta \gamma^2}{\Delta t}\right\rangle + \langle \dot{\gamma}_c\rangle \right) n(\gamma,t)\right]-\frac{n}{t_{\rm esc}}+Q(\gamma,t)\,,
\end{equation}
where $Q\propto \delta(\gamma-\gamma_0)$ with $\gamma_0$ the Lorentz factor of injected particles. For the following calculations we 
fix the Lorentz factor $\gamma_0$ to the value of $100$. The efficiency of shear acceleration depends strongly on the velocity profile of 
the jet, thus in principle the acceleration rate varies with the position in the jet. However, provided the flow speeds are not relativistic and 
the velocity profile is linearly decreasing, an approximate space-averaged spectrum of accelerated particles can be derived. For a homogeneous 
and isotropic spatial distribution of particles, this could be achieved by defining average Fokker--Planck coefficients, i.e.,
\begin{equation}\label{mean_pt}
{\left \langle \frac{\Delta\gamma}{\Delta t}\right \rangle}=\frac{\int 2\pi r \langle \frac{\Delta\gamma}{\Delta t} \rangle (r)dr}{\int 2\pi r dr}\,
\end{equation} and
\begin{equation}\label{mean_ppt}
{\left \langle \frac{\Delta\gamma^2}{\Delta t} \right \rangle}=\frac{\int 2\pi r \langle \frac{\Delta\gamma^2}{\Delta t} \rangle (r)dr}{\int 2\pi r dr}\,.
\end{equation} 
{ In general, the nonthermal particle distribution does not need to be homogeneous and spatial transport may well be non-negligible in realistic 
large-scale jets. While the spectral results derived in the following should thus be interpreted with appropriate care, they provide a useful starting 
point for exploring the potential and characteristics of distributed acceleration processes that, based on jet observations, need to be occurring all along the large-scale jet.}

The interaction between the jet and the ambient medium is naturally expected to lead to some velocity shear at the side boundary of the jet.
Kelvin-Helmholtz instabilities may then drive turbulence which can scatter particles. As an example, we consider a cylindrical jet 
with transverse radius $r_j$, the jet velocity ($z$-component) being constant from $r=0$ to $r_1=0.9r_j$, and decreasing linearly from 
$r_1$ to $r_j$, i.e.,
\begin{equation}
\beta_j=\left \{
\begin{array}{ll}
\beta_{j,0}, & r<0.9r_j,\\
\beta_{j,0}-\frac{\beta_{j,0}}{\Delta L}(r-0.9r_j), & r \geq 0.9r_j.
\end{array}
\right.
\end{equation}
In the presence of turbulence, we generally expect that particle energization by stochastic and shear effects can occur. To study this, the 
Fokker-Planck coefficients in Eq.~(\ref{fp_eq}) are considered to be composed of two parts, i.e. of the relevant terms for shear and 
stochastic acceleration, respectively, that is $\langle \frac{\Delta \gamma^2}{\Delta t}\rangle=\langle \frac{\Delta \gamma^2}{\Delta t}\rangle_{\rm st}
+\langle \frac{\Delta \gamma^2}{\Delta t}\rangle_{\rm sh}$, and $\langle \frac{\Delta \gamma}{\Delta t}\rangle=
\langle \frac{\Delta \gamma}{\Delta t}\rangle_{\rm st}+\langle \frac{\Delta \gamma}{\Delta t}\rangle_{\rm sh}$. 
We employ
\begin{equation}
\langle \frac{\Delta \gamma^2}{\Delta t}\rangle_{\rm st}= \frac{\xi \Gamma_A^4\beta_A^2\gamma^2c}{r_g^{2-q}
\Lambda_{\rm max}^{q-1}}\propto \gamma^{q}
\end{equation}
to incorporate the diffusion coefficient of stochastic acceleration, with $\beta_A$ being the Alfv\'{e}n speed and $\Gamma_A=
1/\sqrt{1-\beta_A^2}$. Given that the system is in detailed balance, the related average rate of change of Lorentz factor (energy) can be 
calculated from the Fokker-Planck relation:
\begin{equation}
\langle \frac{\Delta \gamma}{\Delta t}\rangle_{\rm st}=\frac{1}{2\gamma^2}\frac{\partial}{\partial \gamma}
\left[\gamma^2 \langle \frac{\Delta \gamma^2}{\Delta t}\rangle_{\rm st} \right]\propto \gamma^{q-1}.  
\end{equation}
We then solve Eq.~(\ref{fp_eq}) numerically for the time-dependent particle spectrum based on a finite difference method (see the Appendix for 
details). The results are presented and discussed below.  For comparison and for a cross-check, we also provide asymptotic analytical solutions. 
Intuitively, we expect the spectrum to consist of three parts. At low energies, stochastic acceleration is faster than shear acceleration and hence 
will dominate the formation of the particle energy distribution. Shear acceleration becomes more efficient at higher energies; in this energy 
region it takes over and shapes the energy distribution of particles. Finally, starting at a certain energy, energy losses of radiative character and 
due to particle escape will stop further acceleration, leading to a spectral cutoff.  These effects result in the formation of a rather complex energy 
distribution for the accelerated particles. 

For simplicity, we neglect diffusive escape when deriving the analytic solutions and consider synchrotron radiation as the only cooling mechanism. 
Then, the time-integrated (quasi-steady-state) solution for the spectrum of accelerated particles can be presented in the form
(see the Appendix)
\begin{equation}\label{anal_sol}
n(\gamma)\propto \left \{
\begin{array}{lll}
\gamma^{1-q}  && \gamma_0 < \gamma < \gamma_{\rm eq},\\
\gamma^{q-3}  && \gamma_{\rm eq} \leq \gamma < \gamma_{\rm max},\\
\gamma^{2}{\rm exp}\left[-\frac{6-q}{q-1}\left(\frac{\gamma}{\gamma_{\rm max}}\right)^{q-1} \right] && \gamma_{\rm max} < \gamma.
\end{array}
\right.
\end{equation}
Here, $\gamma_{eq}$ is the Lorentz factor where the stochastic acceleration rate equals the shear acceleration rate:
\begin{equation}
\gamma_{\rm eq}= \,\left[ \frac{15(q+2)}{2(6-q)} \left(\frac{eB}{mc^2}\right)^{4-2q}\frac{\xi^2\Gamma_A^4\beta_A^2c^2}{A^2\Lambda_{\rm max}^{2(q-1)}} 
\right]^{\frac{1}{4-2q}},
\end{equation}
while $\gamma_{\rm max}$ is the Lorentz factor where the shear acceleration rate equals the synchrotron cooling rate. A structure of pile-up followed by a cutoff appears after $\gamma_{\rm meax}$ due to the cooling. The 
(isotropic) synchrotron cooling timescale for a relativistic charged particle is given by $t_{\rm syn}=\frac{9m^3c^5}{4e^4\gamma B^2}$. 
From the condition $t_{\rm  acc,shear}=t_{\rm syn}$, one finds
\begin{equation}\label{gmax}
\gamma_{\rm max}=\left[ \frac{3(6-q)}{20}\Gamma_{j}^4\left(\frac{\beta_{j,0}}{\Delta L}\right)^2\xi^{-1}(mc^2)^{5-q}e^{q-6}B^{q-4}
\Lambda_{\rm max}^{q-1}  \right]^{1/(q-1)}.
\end{equation}
We note that for $q=1$, the energy-dependence of the timescale for shear acceleration is the same as that for synchrotron cooling. Thus, if 
synchrotron cooling is not fast enough to stop the acceleration at low energies where stochastic effects still dominate, it will not be able to do 
so at higher energies either. However, this does not mean that a particle can be accelerated to infinitely high energies. Indeed, there exist at 
least two more constraints on the maximum particle energy. As the energy of a particle increases, its mean free path $\lambda$ also increases 
(except for $q=2$), becoming comparable, at some stage, to the size of the shearing region $\Delta L$. The propagation can then no longer 
be approximated by diffusion and the particle would escape the acceleration region without undergoing much scattering. This constraint (i.e., 
$\lambda<\Delta L$) is somewhat similar to the so-called Hillas condition, which arises from the confinement of particle in the acceleration 
zone, but is more restrictive since the mean free path of a particle in the quasi-linear approximation exceeds its Larmor radius (i.e., $r_g < 
\lambda$). The other constraint is related to the requirement of resonant interactions with MHD waves. The upper limit of the longest wavelength 
$\Lambda_{\rm max}$ is probably comparable to the jet width $r_j$, which implies $r_g<r_j$. Both constraints are related to the particle's 
Larmor radius. Given that $r_g < \lambda$ and $\Delta L \leq r_j$, the constraint from particle confinement is usually more important.

The above considerations show that under the joint operation of stochastic acceleration, shear acceleration, and synchrotron cooling, the 
accelerated particle spectrum contains three segments for $1<q<2$. According to Eq.~(\ref{anal_sol}), the synchrotron spectrum could be
approximated by
\begin{equation}\label{anal_syn}
\nu F_{\nu} \propto \left\{
\begin{array}{lll}
\nu^{(4-q)/2} ~~~~\nu<\nu_{\rm eq}, \\
\nu^{q/2} ~~~~ \nu_{\rm eq}\leq\nu < \nu_{\rm max}, \\
\nu^{4/3}{\rm exp}\left[-(\frac{\nu}{\nu_{\rm max}})^{\frac{q-1}{q+1}} \right], \nu_{\rm max} \leq \nu.
\end{array}
\right.
\end{equation}
where $\nu_{\rm eq}\simeq\frac{1}{4\pi}\frac{\gamma_{\rm eq}^2eB}{m_ec}$ and $\nu_{\rm max}\simeq \frac{1}{4\pi}\frac{\gamma_{\rm max}^2eB}{m_ec}$.
The lowest-energy segment of the spectrum, dominated by electrons from stochastic acceleration, is usually in the radio band. The middle segment 
of the spectrum, in the IR/optical/UV range, is emitted by electrons accelerated by shear acceleration. At the highest energies, likely to be in the X-ray band, the spectrum shows a hardening, which is due to the cooling pileup, coupled with an exponential-like cutoff. We note that in these simple 
realizations the synchrotron spectrum at the highest energies may be decreasing sub-exponentially (e.g., $\propto \exp[-(\nu/\nu_{\rm max})^{1/4}]$ for 
$q=5/3$, also see \citealt{Zirakashvili07}), which would increase the possibility for X-ray detection. Particle escape, on the other 
hand, could lead to some softening of the electron spectrum, resulting in a correspondingly softer synchrotron spectrum where, for example, the possible cooling pileup is smoothed out and the shape of the cutoff is closer to an exponential one. Obviously, the highest energy part is sensitive to details of the assumed escape model and the extent to which the highest energy particles are constrained by the escape. 
%cf. Fig.~\ref{no_esc},

\begin{figure}
\includegraphics[width=0.6\textwidth]{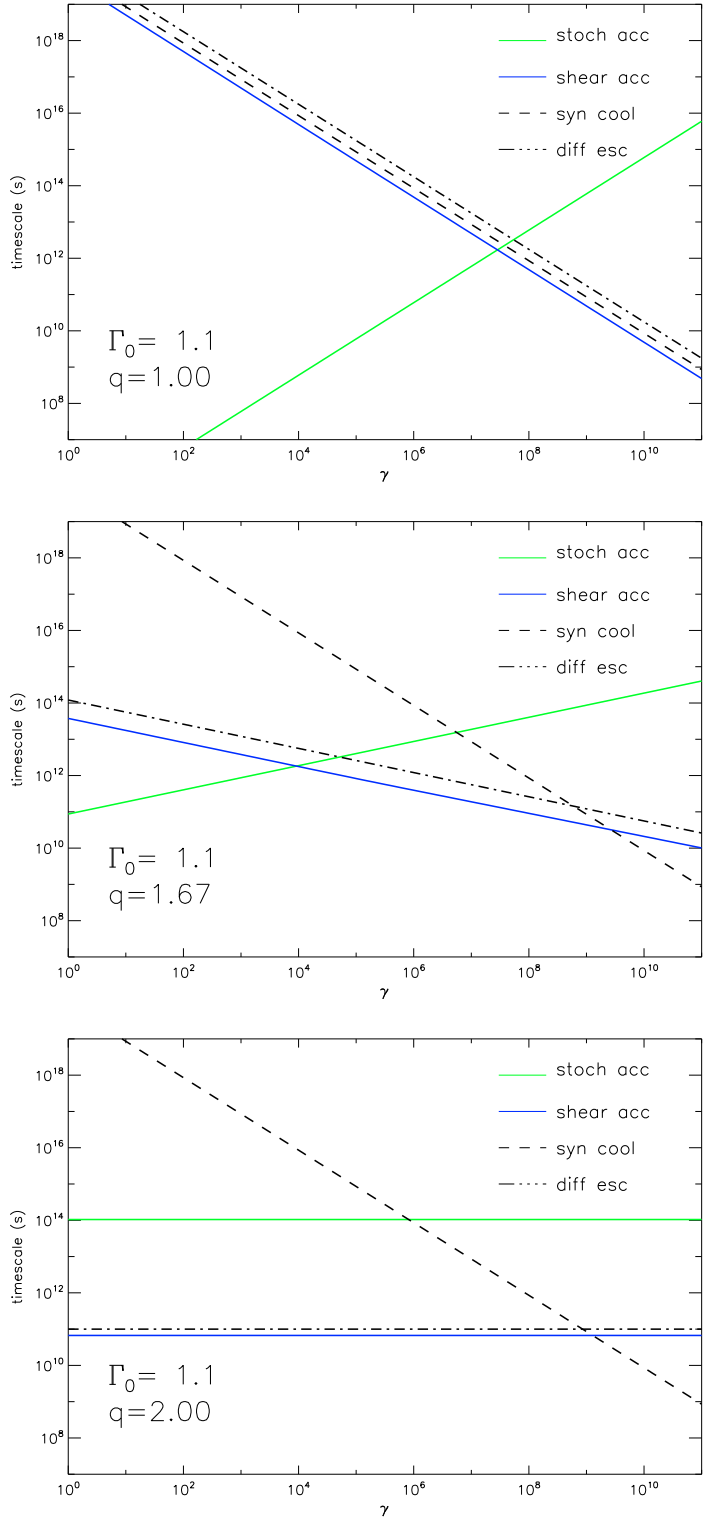} \caption{Timescales of different processes as a function of electron energy (Lorentz factor) for different types of turbulence 
(top: Bohm-type turbulence; middle: Kolmogorov-type turbulence; bottom: hard-sphere approximation). The green solid line shows the timescale for stochastic acceleration 
while the blue solid line show that for shear acceleration. The dashed line represents the timescale for synchrotron cooling. The dash--dotted line is the timescale of diffusive 
escape. Other parameters: $B_0=3\,\mu$G, $\xi=0.1$, $r_j=10^{20}\,$cm, $\Gamma_{j,0}=1.1$, $\Lambda_{\rm max}=10^{18}\,$cm.}\label{timescale}
\end{figure}

Fig.~\ref{timescale} shows the characteristic timescales of electrons for different processes and different types of turbulence.  As one can see, 
for $q=1$ (Bohm-type diffusion regime) and $q=5/3$ (Kolmogorov-type diffusion), stochastic acceleration plays an important role at low energies 
while the shear acceleration dominates at higher energies. The maximum energy achievable due to shear acceleration is about 100 times higher 
than that in stochastic acceleration. It should be noted that around $\gamma\sim 10^9$, the mean free path of electrons becomes comparable 
to the size of the assumed shearing region ($\Delta L=r_j/10=10^{19}\rm cm$), so that some of the electrons will directly escape the jet without 
being scattered again. As mentioned above, in our numerical code the acceleration rate of these electrons is set to zero. This actually leads to a 
rapid rise in the average acceleration timescale beyond $\gamma \sim 10^9$, and is the main cause of the cutoff in the spectrum for 
$q=1$. For $q=2$, on the other hand, the mean free path is independent of particle energy and for the adopted parameters shear acceleration 
is more efficient than stochastic acceleration. These results suggest that, compared to stochastic acceleration, shear acceleration could boost 
maximum electron energies by up to two orders of magnitude, and, correspondingly, the maximum energy of synchrotron radiation by up to four 
orders of magnitude.

\begin{figure}
\includegraphics[width=0.6\textwidth]{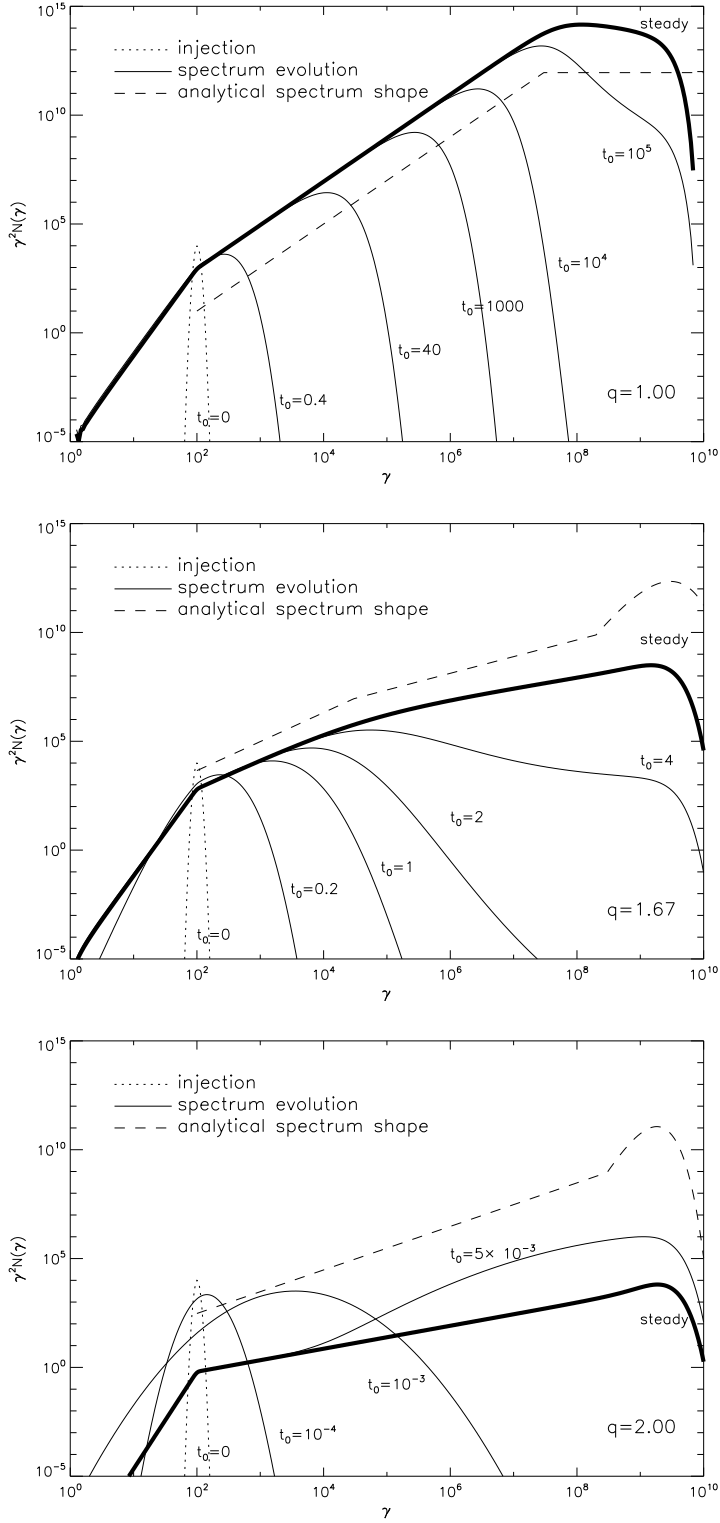} \caption{The evolution of the electron spectrum with time for different types of turbulence (top: Bohm-type turbulence; middle: 
Kolmogorov-type turbulence; bottom: hard-sphere approximation). The thick solid lines show the spectrum in steady state while the thin solid lines are the spectrum at 
different times after injection, as marked in the figure. {  Dashed lines show the time-integrated (quasi steady-state) spectral shape of the analytic solutions.} All the adopted 
parameters are the same as those in Fig.~\ref{timescale}}\label{spec}
\end{figure}

\begin{figure}
\includegraphics[width=0.6\textwidth]{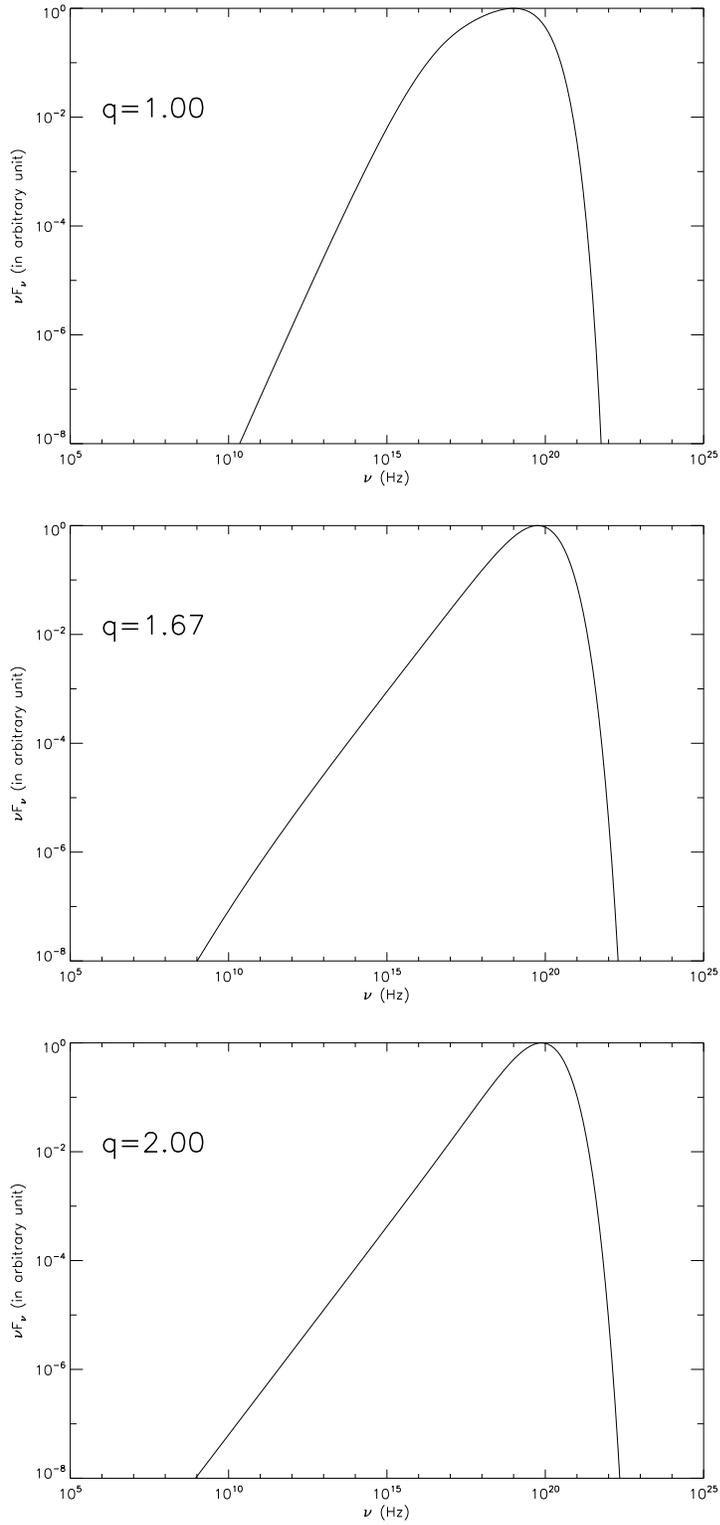} \caption{The spectrum of synchrotron emissions by the accelerated electrons in steady state for different types of turbulence 
(top: Bohm-type turbulence; middle: Kolmogorov-type turbulence; bottom: hard-sphere approximation). All the adopted parameters are the same as those in 
Fig.~\ref{timescale}}\label{syn_spec}
\end{figure}

Fig.~\ref{spec} exhibits the numerical results for the time evolution of the accelerated electron spectrum for $q=1$, $q=5/3$ and $q=2$. 
The thick solid lines represent the spectrum in a steady state, while thin solid lines show the spectrum at different times before reaching the steady state, with the time being expressed in units of $t_0$ which is related to the timescale for stochastic acceleration of the particles at injection, i.e. $t_{0}=
\gamma_0^2/\langle \frac{\Delta \gamma^2}{\Delta t} \rangle_{\rm st}(\gamma=\gamma_0)$. 
The dashed lines are the profiles of the analytical solutions in the (quasi-)steady state without considering escape. We can see that the shapes of 
the numerical solutions generally follow that of the analytical ones, except that (1) the spectral slopes of the part dominated by shear acceleration 
are slightly softer than the analytical ones and  (2) the pileup bump feature is not as pronounced or even vanishes for the chosen parameters. Both 
differences are due to the neglecting the particle escape in the analytical treatment. And such differences in the electron spectrum are also reflected in the synchrotron spectrum, as illustrated in Fig.~\ref{syn_spec}. Although the peak of the $\nu F_\nu$ spectrum in the chosen example can still be as high 
as $10^{19}-10^{20}$ Hz, the high-energy spectral indices are softer and the cutoff are sharper compared to those depicted by the analytical 
synchrotron spectrum (Eq.\ref{anal_syn}). 

\begin{figure}
\includegraphics[width=0.6\textwidth]{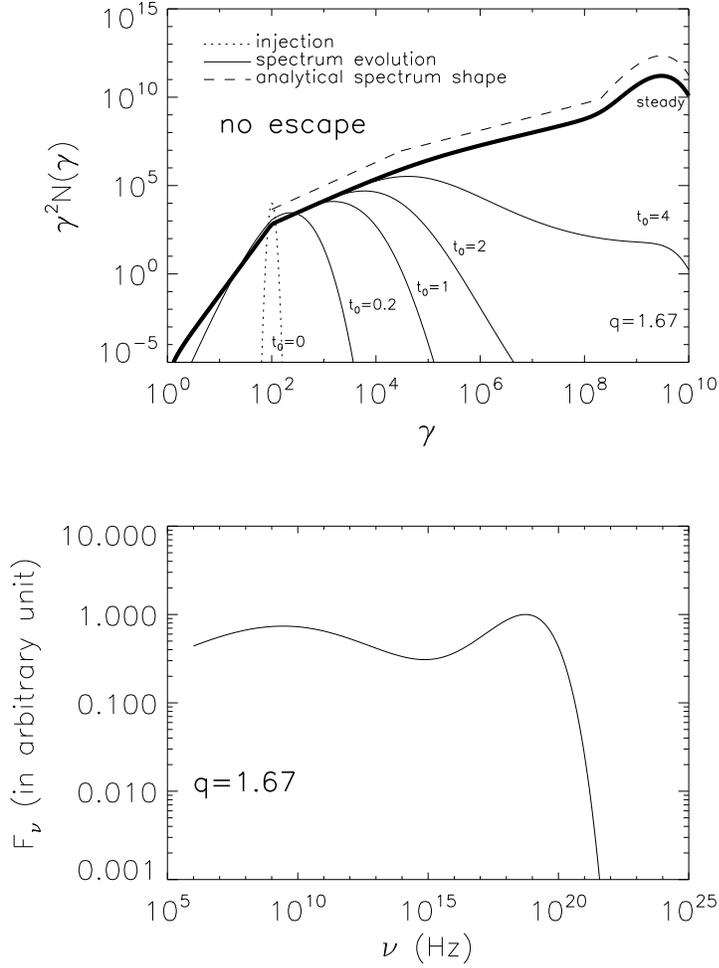} \caption{The evolution of the electron spectrum with time and associated synchrotron radiation $F_{\nu}$ in a steady state for $q=5/3$, 
assuming inefficient escape. {The dashed line shows the time-integrated (quasi-steady-state) spectral shape of the analytic 
solutions.}\label{no_esc}}
\end{figure}

To further illustrate the possible effect of escape, the spectral evolution of the electron distribution for $q=5/3$ and the synchrotron spectrum in a steady 
state are shown in Fig.~\ref{no_esc}, in 
which all parameters are kept the same as in Fig.~\ref{spec}, but where the diffusive escape term $n/t_{\rm esc}$ in Eq.~(\ref{fp_eq}) 
has been {artificially reduced by a factor of 1000} and the effect of the direct escape on the Fokker--Planck coefficients at high energies is ignored. 
As we can see in the upper panel of Fig.~\ref{no_esc}, the slope of the electron spectrum at high energies is well consistent with the analytical one. 
We can also see the change in the synchrotron spectral indices is also evident in the lower panel, where the y-axis is set to be $F_\nu$ instead of 
$\nu F_\nu$ as in the other figures.

\begin{figure}
\includegraphics[width=0.6\textwidth]{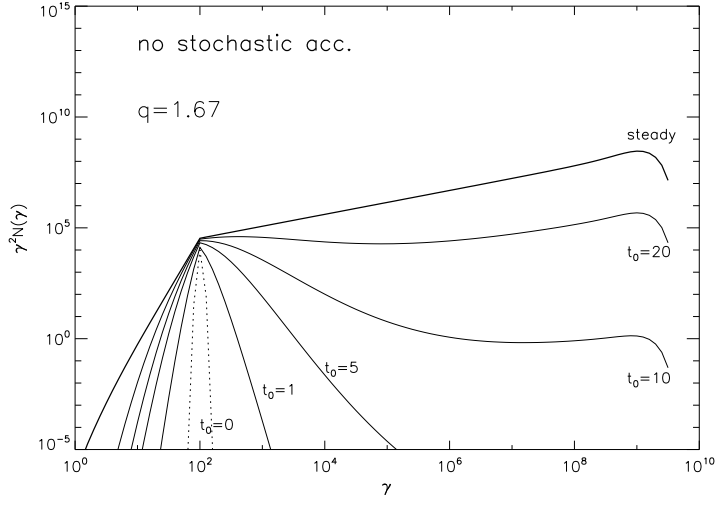} \caption{The evolution of the electron spectrum for $q=5/3$ where stochastic acceleration has been ignored.\label{no_sto}}
\end{figure}

Although stochastic acceleration does not affect the high-energy spectrum, it plays an important role in the whole acceleration process. 
Fig.~\ref{no_sto} shows an example of the time evolution of the electron spectrum, where the terms related to stochastic acceleration 
in Eq.~(\ref{fp_eq}) have been removed while all parameters are kept the same as in the middle panel of Fig.~\ref{spec}. We can see 
that the time required to reach the steady state has increased by about a factor of five. Such an increment will be further amplified, if the Lorentz
factor of particle injection, $\gamma_0$,  is lower or the spectral index of the turbulence $q$ is smaller. 
Note that if we assume that the jet extends to a length of 100\,kpc, the dynamical time scale associated with the inner part of the shear would 
be $\sim 10^{13}$\,s. For the adopted parameters, shear acceleration may then not be able to efficiently accelerate from energies much 
below $\gamma \sim 100$. Stochastic acceleration thus serves as an important pre-acceleration mechanism when the energy of the injected 
electrons is low, and thereby boosts the whole acceleration process.

We note that in the calculations presented, we have focused only on the distribution of particles in momentum (Lorentz factor) space and neglected 
their spatial distribution. Given the assumption that the flow is not highly relativistic and that all other parameters are homogeneous over the whole 
region, a consideration of the spatial distribution may not much affect the general momentum distribution, although one might expect that the maximum 
energy would be constrained somewhat further. While in our current treatment the escape of particles is related only to their energies, in reality their 
position should also affect the escape probability, e.g. particles near the boundary of the acceleration region should have a higher chance of escape 
than those closer to the center. Hence achievable maximum energies are likely to be somewhat smaller than the one obtained above if 
shear acceleration is not limited by cooling.

\section{Potential for the acceleration of ltrahigh-energy cosmic rays (UHECRs)}
Relativistic AGN jets have long been suggested as candidate sites for the origin of UHECRs. 
In particular, the properties of the hotspots, the knots, the lobes, and the extended jet itself do not only satisfy the requirement for the confinement 
of these energetic particles (i.e., the so-called Hillas criterion), but may also satisfy the stricter constraint concerning the required source's energy 
budget that arises due to the cooling processes of particles \citep{Aharonian02b}. Several scenarios have been considered for the acceleration 
of UHECRs in AGN jets, such as acceleration at the jet termination shock \citep[e.g.,][]{Biermann87, Rachen93}, { a single $\Gamma^2$-kick 
up of Galactic-like cosmic rays in blazar-like jets with $\Gamma>30$} \citep[e.g.,][]{Biermann09,GK10, Caprioli15}, pinch acceleration 
in a jet with electric current \citep[e.g.,][]{Vlasov90, Berezinsky06}, and shear acceleration in a shearing jet \citep[e.g.,][]{Ostrowski98,Rieger07}. 
In this section, our main purpose is to explore the maximum achievable proton energy under the operation of shear acceleration in a 
mildly relativistic jet.

\begin{figure} 
\includegraphics[width=0.95\textwidth]{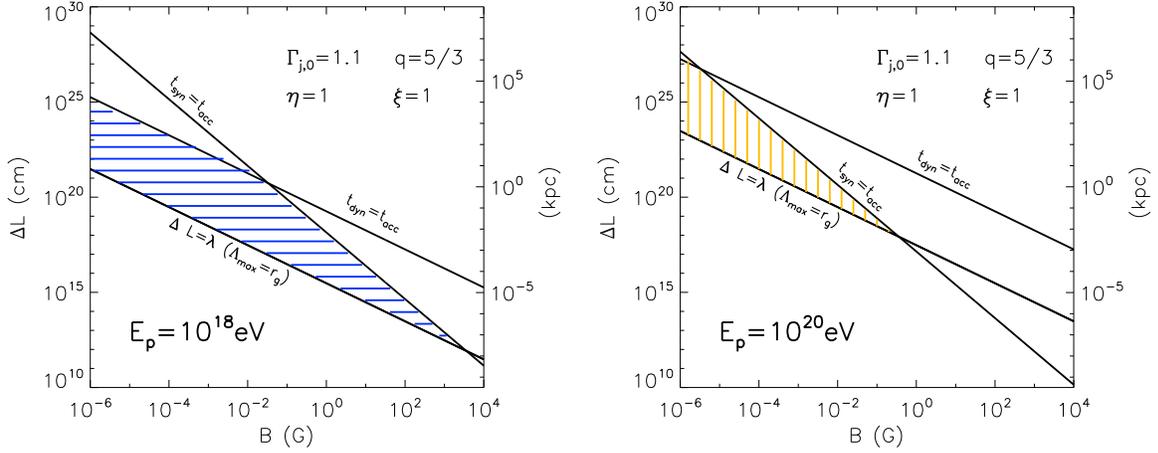} \caption{Parameter space for the acceleration of a $10^{18}\,$eV proton (left panel, blue hatched region) 
and a $10^{20}\,$eV proton (right panel, orange hatched region) in a shearing jet with a bulk Lorentz factor of $\Gamma_{j,0}$ in the 
spine. A Kolmogorov-type ($q=5/3$) turbulence is assumed.\label{p_acc}}
\end{figure}

As protons are more massive, they are less affected by synchrotron cooling than electrons of the same energy. The most restrictive 
constraint on the maximum proton energy in shear  acceleration most likely comes from particle confinement, which depends on the size of 
the acceleration region and the mean free path of the proton. One may thus expect that larger magnetic fields and larger sizes of the 
acceleration regions will be conducive to obtaining higher proton energies. 
Consider as an example the case $q=5/3$ for the linearly decreasing profile of jet velocity employed above, and assume that the longest 
interacting wavelength $\Lambda_{\rm max}=\eta\Delta L$, $\eta\leq 1$. Then, for a jet to be able to produce protons above a certain energy 
$E$, the following conditions should be satisfied:\\
(1) Protons should be confined in the shearing region, i.e., $\Delta L > \lambda=\xi^{-1}r_g^{2-q}\Lambda_{\rm max}^{q-1}$ 
which translates to 
\begin{equation}
\Delta L > 3\times 10^{20} \xi^{-3} \eta^{2} \left(\frac{E}{10^{18}\rm eV}\right)\left(\frac{B}{10\mu\rm G}\right)^{-1}\ \rm cm\,.
\end{equation} 
(2) The longest interacting turbulent wavelength should be larger than the Larmor radius of the proton to ensure resonance 
between wave and particle, and hence allow for acceleration. This requires $\Lambda_{\rm max}> r_g$, or
\begin{equation}
\Delta L > 3\times 10^{20} \eta^{-1} \left(\frac{E}{10^{18}\rm eV}\right)\left(\frac{B}{10\mu \rm G} \right)^{-1}
\end{equation}
Note that this gives a similar dependence on $E$ and $B$ as obtained by the first requirement.\\
(3) The timescale for acceleration should be less than the synchrotron cooling timescale, $t_{\rm acc} < t_{\rm  syn}$, which 
gives 
\begin{equation}
\Delta L < 8 \times 10^{26} \eta^{1/2} \xi^{-3/4}  \left(\frac{E}{10^{18}\rm eV}\right)^{-1/2}\left(\frac{B}{10\mu \rm G}\right)^{-7/4}
\left(\frac{\Gamma_{j,0}}{1.1}\right)^3\left(\frac{\beta_{j,0}}{0.42}\right)^{3/2} \rm cm\,.
\end{equation} 
(4) The time scale for acceleration should be less than the advective timescale or the dynamical timescale ($ t_{\rm acc}< 
t_{\rm dyn}\sim c~d/\beta_{j,0}$). 
Assuming that the jet length $d$ is related to the width of the shearing boundary by $\Delta L = \rho_w d$ (typically $\rho_w \leq 0.1$), 
we find 
\begin{equation}
\Delta L < 1.8 \times 10^{24} \eta^{2}\xi^{-3} \left(\frac{\rho_w}{0.01}\right)^{-3}\left(\frac{\Gamma_{j,0}}{1.1}\right)^{12}\left(\frac{\beta_{j,0}}{0.42}\right)^{3}
\left(\frac{E}{10^{18}\rm eV}\right)\left(\frac{B}{10\mu\rm G}\right)^{-1}\ \rm cm
%\Delta^{2-q} < \frac{6-q}{15}\rho_w^{-1}\Gamma_{j,0}^4\beta_{j,0}\xi^{-1}E^{2-q}e^{q-2}B^{q-2}\eta^{q-1}
\end{equation}
In order to illustrate the permitted parameter space for the width of the shearing boundary $\Delta L$ and the average jet magnetic field $B$, 
Fig.~\ref{p_acc} shows the available parameter space with regard to efficient shear acceleration of a $10^{18}\,$eV proton and a $10^{20}\,$eV 
one, where $\Gamma_{j,0}=1.1$ ($\beta_{j,0}=0.42$), $q=5/3$, $\eta \sim 1$, $\xi \sim 1$ and $\rho_{w}\sim 0.01$ have been 
employed. One can see that $10^{18}$ eV protons may be achievable in jets with relatively plausible parameters such as 10 kpc--1 Mpc in length 
and (1--100) $\mu$Gauss in magnetic field, while conditions for efficient shear acceleration of protons to $10^{20}$ eV in mildly relativistic 
large-scale flows would be quite restrictive and may actually require relativistic flow speeds \citep[e.g.][]{Ostrowski00,Rieger04,Rieger16}. 
Note that the permitted parameter space is generally sensitive to the considered setup and the chosen description of turbulence , and may vary 
significantly depending on individual source characteristics.
{While more detailed and source-specific studies would be needed before firm conclusions can be drawn, the results suggest that shear 
acceleration in large-scale AGN jets could potentially also contribute to the energization of UHECRs.}

\section{Discussion and Conclusion}
%\subsection{On the origin of extended emission in large-scale AGN jets}
The short lifetime of synchrotron X-ray-emitting electrons in AGN jets calls for a spatially distributed mechanism that facilitates their re-energization 
along the kpc--Mpc jets. As shown in this work, stochastic second-order Fermi and shear acceleration processes could provide a plausible means for 
the appearance of UHE electrons throughout the entire large-scale jet. Unlike the IC/CMB model, the jet in this scenario is not required to 
remain highly relativistic at kpc--Mpc scale; a mildly relativistic jet speed would be sufficient to efficiently accelerate electrons and to account for X-ray 
emission via synchrotron radiation. In addition, the combined operation of stochastic and shear acceleration could naturally result in different spectral 
indices at different energy bands. This could make it possible to simultaneously account for the softening in the optical/UV band and the hardening in 
the X-ray band as seen in some sources \citep[see e.g.,][for a comprehensive review of X-ray jets in AGNs]{Harris06}.  Alternatively, acceleration at a 
boundary shear could be solely responsible for the X-ray emission, while the lower-energy radio--optical emission could arise from some other 
mechanisms operating in the central part of the jet \citep[e.g.,][]{Jester06}. In this scenario, one might expect the transverse X-ray profile of the jet 
to show a limb-brightening feature since the acceleration would preferentially happen at the surface of the jet. This may, however, be difficult to probe 
given the angular resolution of current instruments. 

%\subsection{Generation of magnetohydrodynic (MHD) turbulence in a shearing flow}
Essential to the current analysis is the assumption that sufficient MHD turbulence is present within the jet and that the flow is capable of maintaining a 
transverse velocity shear. Shearing plasmas are indeed known to drive the generation of MHD turbulence due to, e.g. streaming, Kelvin-Helmholtz  or Mushroom
instabilities and a variety of different explorations have been conducted in this regard \citep[e.g.][]{Gruzinov08, Zhang09, Alves12, Liang13a,Liang13b, 
Alves14, Nishikawa14, Alves15}. In reality the situation may be rather complex. In particular, the velocity shear could be smeared out as a result of particle 
acceleration, dissipation into heat, and generation or amplification of magnetic fields, while some transverse structure could be generated due to the 
mixing of plasma in different layers of the jet. The feedback of the generation of MHD turbulence on the velocity profile is likely to influence the 
acceleration process. Viscous effects related to particle acceleration, on the other hand, are probably not able to introduce a significant modification 
of the flow \citep{Rieger06}. Depending on the plasma properties, particles might also be accelerated in the parallel direction of the bulk plasma flow 
due to the generation of small-scale electric fields \citep[e.g.,][]{Alves14,Nishikawa14}. Incorporation of such effects, however, is beyond the treatment 
presented here.

Detailed studies of the diffusion properties in shearing plasmas will be important to better determine the efficiency of shear acceleration. In the present 
work, we have considered isotropic spatial diffusion of particles for simplicity. {Numerical simulations would be needed to better assess and qualify 
this assumption.} We note that the overall magnetic field configuration is important since it affects the degree of spatial diffusion. In the absence of strong 
turbulence, for example, a dominant longitudinal magnetic field would lead to a suppression of transverse diffusion and decrease the difference in bulk flow velocity that a particle could sample within one mean free path, making shear acceleration less efficient.
With a focus on large-scale AGN jets, we have restricted our discussion to mildly relativistic jet speeds, where the difference in flow velocity  experienced by an electron within one mean free path is nonrelativistic. The energy increment per scattering is then much smaller than the particle's energy so that 
a Fokker--Planck approach becomes feasible. Once the energy gain becomes relativistic, a more extended treatment would be needed, including an
incorporation of the anisotropy in the scattering of particles. 

To summarize, {our results suggest that in the case of AGNs} shear acceleration along with stochastic second-order Fermi acceleration could provide 
favorable conditions for the presence of UHE electrons throughout the entire large-scale jet. 
Based on a heuristic microscopic analysis we have shown here that energetic particles can experience efficient acceleration in a mildly relativistic 
turbulent shearing jet. Assuming isotropic diffusion of charged particles, the average acceleration and the momentum diffusion rate due to the shear 
mechanism have been evaluated. In general, turbulence will also lead to some stochastic second-order Fermi acceleration. In order to study the possible
interplay between systematic (gradual) shear and stochastic second-order Fermi effects, a simple Fokker--Planck approach has been employed. By solving the 
corresponding Fokker--Planck equation including synchrotron cooling and escape of electrons, the time-dependent evolution of the particle distribution in 
momentum space has been obtained and compared to analytical (steady-state) solutions. The results illustrate that although shear acceleration is conducive 
to the acceleration of high-energy particles, it is usually not efficient for accelerating low-energy particles. In this regard, stochastic acceleration can play an 
important role as a pre-acceleration process and provide the required high-energy seed particles for shear acceleration, allowing for significant boosting 
of the whole process. The resultant particle distributions are expected to exhibit multiple components characterized by different (turbulence-dependent) 
spectral indices. Examples to this end are presented for typical parameters applicable to AGN jets. The results indicate the potential of the considered 
processes and motivate further in-depth studies with regard to the extended X-ray emission in AGN jets and the origin of UHECRs.

\acknowledgements FMR kindly acknowledges support by a DFG Heisenberg Fellowship (RI 1187/4-1).

\appendix
\section{Solutions to the Fokker--Planck Equation}
\subsection{Analytical solution assuming quasi-steady state}
%Let's write the Eq.~(\ref{fp_eq}) here again for convenience. 
%\begin{equation}
%\frac{\partial n(\gamma,t)}{\partial t}=\frac{1}{2}\frac{\partial}{\partial \gamma}\left[\left \langle \frac{\Delta \gamma^2}{\Delta t}\right\rangle\frac{\partial n(\gamma,t)}{\partial \gamma} \right]- \frac{\partial}{\partial \gamma}\left[ \left( \left\langle \frac{\Delta \gamma}{\Delta t}\right\rangle-\frac{1}{2}\frac{\partial}{\partial \gamma}\left\langle \frac{\Delta \gamma^2}{\Delta t}\right\rangle + \langle \dot{\gamma}_c\rangle \right) n(\gamma,t)\right]-\frac{n(\gamma, t)}{t_{\rm esc}}+Q(\gamma,t)
%\end{equation}
Considering a quasi-steady state in the energy range of interest ($\partial n/\partial t=0$), neglecting the escape term\footnote{In principle, a full steady state cannot 
be reached for the entire system in the presence of injection and without escape. However, a steady state may still be achievable above a certain threshold energy. 
Particle injection could be balanced by a particle flux passing this threshold toward lower energies, though this may require some specific boundary conditions at low energies (for a discussion of this and related issues, see e.g. Stawarz \& Petrosian 2008). Here we are mainly concerned with an analytical approximation of the spectral shape at high energies for reasons of comparison, leaving aside details on boundary conditions and the behavior of the spectral shape at low energy.}, and 
keeping in 
mind the relation between the two Fokker--Planck coefficients, 
Eq.~(\ref{fp_eq}) can be reduced to
\begin{equation}\label{ss}
\frac{1}{2}\frac{\partial}{\partial\gamma}\left[\left\langle \frac{\Delta \gamma^2}{\Delta t} \right\rangle \frac{\partial n}{\partial \gamma} \right] - \frac{\partial}{\partial\gamma}\left[\left(\frac{1}{\gamma}\left\langle \frac{\Delta \gamma^2}{\Delta t} \right\rangle + \langle \dot{\gamma}_c\rangle \right) n\right] + Q(\gamma, t)=0
\end{equation}
Here $\left\langle \frac{\Delta \gamma^2}{\Delta t} \right\rangle=\left\langle \frac{\Delta \gamma^2}{\Delta t} \right\rangle_{\rm st}+\left\langle \frac{\Delta \gamma^2}{\Delta t} \right\rangle_{\rm sh}$; it includes the term for both the shear acceleration and stochastic acceleration. We further consider the injection function $Q(\gamma,t)=C\delta(\gamma-\gamma_0)$ with $C$ being a constant. For $\gamma < \gamma_{\rm max}$, the cooling term is negligible and we have
\begin{equation}\label{reduced}
\frac{1}{2}\frac{\partial}{\partial\gamma}\left[\left\langle \frac{\Delta \gamma^2}{\Delta t} \right\rangle \frac{\partial n(\gamma)}{\partial \gamma}\right]-\frac{\partial}{\partial\gamma}\left[\frac{1}{\gamma}\left\langle \frac{\Delta \gamma^2}{\Delta t} \right\rangle n(\gamma)\right] + Q(\gamma, t)=0
\end{equation}
Then, for $\gamma < \gamma_0$, integrating this equation from 1 to $\gamma$, and taking into account that we have $n(\gamma=1)\to 0$, $\partial n(\gamma=1, t)/\partial \gamma \to 0$, we obtain 
\begin{equation}\label{gamma0-}
\frac{n}{\gamma}-\frac{1}{2}\frac{\partial n}{\partial \gamma}=0,
\end{equation}
which results in $n(\gamma)=  C_0\gamma^2$ with $C_0$ being a normalization factor. Note that Eq.~(\ref{fp_eq}) is converted from Eq.~(\ref{fp_eq_p}) with the assumption 
$p=\gamma mc$ valid for $\gamma \gg 1$. The solution without such a substitution would be $n\propto p^2 \propto (\gamma-1)^2$. 

For $\gamma_0<\gamma<\gamma_{\rm eq}$, we integrate Eq.~(\ref{reduced}) from $\gamma_0+0^+$ to a certain $\gamma$: 
\begin{equation}\label{gamma0+}
\left(\frac{1}{2}\frac{\partial n}{\partial \gamma}-\frac{n}{\gamma}\right)\left\langle \frac{\Delta \gamma^2}{\Delta t} \right\rangle-C_1=0
\end{equation}
where $C_1=\left[\left(\frac{1}{2}\frac{\partial n}{\partial \gamma}-\frac{n}{\gamma}\right)\left\langle \frac{\Delta \gamma^2}{\Delta t} \right\rangle\right]_{\gamma=\gamma_0+0^+}$ is a constant. Given that stochastic acceleration dominates in 
this energy range, we have $\left\langle \frac{\Delta \gamma^2}{\Delta t} \right\rangle=\left\langle \frac{\Delta \gamma^2}{\Delta t} \right\rangle_{\rm st}\propto \gamma^q$. Then 
we can find the solution of the above equation as
$n(\gamma)=\gamma^2\left[ K - \frac{2C_1\gamma^{-1}}{(1+q)\left\langle \frac{\Delta \gamma^2}{\Delta t} \right\rangle}\right]$.
Here $K$ is an integration constant which should be 0, otherwise the spectrum becomes $N(\gamma)\propto \gamma^2$ at high energies, independent on the diffusion 
coefficient, which, however, does not have a physical meaning. Thus, we have 
\begin{equation}
n(\gamma)=-\frac{2C_1\gamma}{(1+q)\left\langle\frac{\Delta \gamma^2}{\Delta t}\right\rangle_{\rm st}}\propto \gamma^{1-q},  ~~~~ \gamma_0 < \gamma <\gamma_{\rm eq}.
\end{equation}
Now let us determine the constant $C_1$. Integrating Eq.~(\ref{reduced}) from $\gamma_0+0^-$ to $\gamma_0+0^+$, we find
\begin{equation}
\left[\left(\frac{1}{2}\frac{\partial n}{\partial \gamma}-\frac{n}{\gamma}\right)\left\langle\frac{\Delta \gamma^2}{\Delta t}\right\rangle\right]_{\gamma=\gamma_0+0^+}-\left[\left(\frac{1}{2}\frac{\partial n}{\partial \gamma}-\frac{n}{\gamma}\right)\left\langle\frac{\Delta \gamma^2}{\Delta t}\right\rangle\right]_{\gamma=\gamma_0+0^-}+C=0.
\end{equation}
Note that $\left[\frac{1}{2}\frac{\partial n}{\partial \gamma}-\frac{n}{\gamma}\right]_{\gamma=\gamma_0+0^-}=0$ (see Eq.~(\ref{gamma0-})) and $\left[\left(\frac{1}{2}\frac{\partial n}{\partial \gamma}-\frac{n}{\gamma}\right)\left\langle\frac{\Delta \gamma^2}{\Delta t}\right\rangle\right]_{\gamma=\gamma_0+0^+}=C_1$, and we obtain $C_1=-C$. In addition, 
$N(\gamma)$ should be continuous at $\gamma_0$, i.e., $n(\gamma_0+0^-)=n(\gamma_0+0^+)$, based on which we can obtain the value of $C_0$. 

For $\gamma_{\rm eq}<\gamma<\gamma_{\rm max}$, the cooling term can still be ignored and we integrate Eq.~(\ref{reduced}) from $\gamma_{\rm eq}$ to a certain 
$\gamma$, resulting in
\begin{equation}
\left(\frac{1}{2}\frac{\partial n}{\partial \gamma}-\frac{n}{\gamma}\right)\left\langle \frac{\Delta \gamma^2}{\Delta t} \right\rangle-C_2=0
\end{equation}
with $C_2=\left[\left(\frac{1}{2}\frac{\partial n}{\partial \gamma}-\frac{n}{\gamma}\right)\left\langle \frac{\Delta \gamma^2}{\Delta t} \right\rangle\right]_{\gamma=\gamma_{\rm eq}}$. Note that in this energy range shear acceleration dominates 
over stochastic acceleration so that $\langle\frac{\Delta \gamma^2}{\Delta t}\rangle=\langle\frac{\Delta \gamma^2}{\Delta t}\rangle_{\rm sh}\propto \gamma^{4-q}$. Similarly, 
we can get
\begin{equation}
n(\gamma)=-\frac{2C_2\gamma}{(5-q)\left\langle\frac{\Delta \gamma^2}{\Delta t}\right\rangle_{\rm sh}}\propto \gamma^{q-3}, ~~~~ \gamma_{\rm eq}<\gamma<\gamma_{\rm max}.
\end{equation}

For $\gamma_{\rm max}<\gamma$, the cooling term becomes important. The integration of Eq.~(\ref{ss}) from $\gamma$ to $+\infty$ gives
\begin{equation}
0-\left[\frac{1}{2}\left\langle \frac{\Delta \gamma^2}{\Delta t} \right\rangle \frac{\partial n}{\partial \gamma} -\left(\frac{1}{\gamma}\left\langle \frac{\Delta \gamma^2}{\Delta t} \right\rangle + \langle \dot{\gamma}_c\rangle \right) n \right]=0
\end{equation}
Note that we already substituted $n(\gamma\to \infty)=0$ and $\frac{\partial n}{\partial \gamma}(\gamma \to \infty)=0$ into the above equation. This gives us 
$n(\gamma)=C_3\gamma^2 {\rm exp}\left(\int \frac{2\langle \dot{\gamma}_c \rangle}{\left\langle \frac{\Delta \gamma^2}{\Delta t} \right\rangle}d\gamma\right)$
where $C_3$ is a constant which could be found from the relation $N(\gamma_{\rm max}+0^-)=N(\gamma_{\rm max}+0^+)$. Note that the diffusion term is dominated by shear acceleration. Considering synchrotron cooling with $<\dot{\gamma}_c>\propto -\gamma^2$, one obtains
\begin{equation}
n(\gamma)=C_3\gamma^2{\rm exp}\left[-\frac{6-q}{q-1}\left(\frac{\gamma}{\gamma_{\rm max}}\right)^{q-1} \right], ~~~~ \gamma_{\rm max}<\gamma.
\end{equation}

\subsection{Numerical solution for time--dependent equation}
To obtain the time-dependent solution, we write  Eq.~(\ref{fp_eq}) to a central difference equation as
\begin{equation}
\dot{N}=\phi(\gamma) N" + p(\gamma)N'+q(\gamma)N  + Q 
\end{equation}
with
\begin{align}
& N=n(\gamma ,t),\\
& N'=\frac{n(\gamma+\Delta \gamma,t)-n(\gamma-\Delta \gamma,t)}{2\Delta \gamma},\\
& N"=\frac{n(\gamma+\Delta \gamma,t)-2n(\gamma,t)+n(\gamma-\Delta \gamma,t)}{\Delta \gamma^2},\\
& \dot{N}=\frac{n(\gamma, t+\Delta t)-n(\gamma,t)}{\Delta t}.
\end{align}
and
\begin{align}
& \phi(\gamma)=\frac{1}{2}\left\langle \frac{\Delta\gamma^2}{\Delta t} \right\rangle,\\
& p(\gamma)=\left(\frac{1}{2}\frac{\partial}{\partial \gamma}\left\langle \frac{\Delta\gamma^2}{\Delta t} \right\rangle  -\frac{1}{\gamma}\left\langle \frac{\Delta\gamma^2}{\Delta t} \right\rangle -\langle \dot{\gamma}_c \rangle   \right),\\
& q(\gamma)=\left(\frac{1}{\gamma^2}
\left\langle \frac{\Delta \gamma^2}{\Delta t} \right\rangle -\frac{1}{\gamma}\frac{\partial}{\partial \gamma}\left\langle \frac{\Delta \gamma^2}{\Delta t} \right\rangle + \frac{\partial \langle \dot{\gamma}_c\rangle}{\partial \gamma} -\frac{1}{t_{\rm esc}}  \right).
\end{align}
Since the energy span is very large and we are interested in the spectrum over the entire energy range, there will be too many grid points in energy dimension if we divide it linearly, 
making it extremely time--consuming. So we introduce $x=\ln \gamma$ and $\Delta x=\ln (\gamma+\Delta \gamma) - \ln \gamma$, and redefine
\begin{align}
& N'=\frac{n(\gamma+\Delta \gamma,t)-n(\gamma-\Delta \gamma,t)}{2\Delta x},\\
& N"=\frac{n(\gamma+\Delta \gamma,t)-2n(\gamma,t)+n(\gamma-\Delta \gamma,t)}{\Delta x^2}.
\end{align}
Then, we can divide the energy range logarithmically and find
\begin{equation}\label{num_sol}
N(\gamma, t+ \Delta t)=\left(\frac{\phi(\gamma)}{\gamma^2} (N"-N')+\frac{p(\gamma)}{\gamma}N'+q(\gamma)N +Q(\gamma)\right) \Delta t + N(\gamma, t)
\end{equation}
with the initial condition $N(\gamma, 0)=Q(\gamma)$ and the boundary condition $N(0,t)=N(10^{12},t)=0$. In the analytical treatment, we assume a Dirac function for the 
injection term $Q$. Here, for the injection term we employ a narrow Gaussian function in logarithmic energy space, i.e., $Q(\gamma)={\rm exp}(-100(\ln\gamma-\ln \gamma_0)^2)$. 
Note that the solution depends weakly on the width of the Gaussian function (as long as it is narrow enough) since the broadening will be dominated by the diffusion. Based 
on these conditions and Eq.~(\ref{num_sol}), we employ a Fortran implementation to calculate the spectrum from the beginning of the injection to the steady state.

% Figure 1 (sketch.eps)

% Figure 2 (acc_rate.eps)

%% Figure 3 (spec_q1.eps)
%\begin{figure}
%\includegraphics[width=0.6\textwidth]{fig3} \caption{Timescales for electrons for different processes (top), the evolution of the electron spectrum $\gamma^2 n(\gamma)$ (middle)
%and the resultant (steady-state) synchrotron radiation $\nu F_{\nu}$ (bottom) for the (Bohm-type turbulence) case $q=1$. Other parameters: $B=3\,\mu$G, $\xi=0.1$, $r_j=10^{20}\,$cm, $\Gamma_{j,0}=1.1$, $\Lambda_{\rm max}=10^{18}\,$cm. \label{spec_beta1}}
%\end{figure}
%
%% Figure 4 (spec_q167.eps)
%\begin{figure}
%\includegraphics[width=0.6\textwidth]{fig4} \caption{The same as Fig.~\ref{spec_beta1} but for the (Kolmogorov-type turbulence) case 
%$q=5/3$.\label{spec_beta2}}
%\end{figure}
%
%% Figure 5 (spec_q2.eps)
%\begin{figure}
%\includegraphics[width=0.6\textwidth]{fig5} \caption{The same as Fig.~\ref{spec_beta1} but for the case $q=2$.\label{spec_beta3}}
%\end{figure}

%Figure 6 (spec_q167_noesc.eps)

% Figure 7 (spec_q167_nosto.eps)

% Figure 8 (constraint_pacc1)

\end{document}